\documentclass[aip,apl,reprint,amsmath,amssymb,superscriptaddress]{revtex4-1}
\usepackage{graphicx}
\usepackage{dcolumn}
\usepackage{bm}
\usepackage{textcomp}

\newcommand{\variable}[2]{\ensuremath{#1_{\mathrm{#2}}}}

\begin{document}

\title{Double quantum dot with tunable coupling in an enhancement-mode silicon metal-oxide semiconductor device with lateral geometry}

\author{L. A. Tracy}
\email{latracy@sandia.gov}
\affiliation{Sandia National Laboratories, Albuquerque, New Mexico 87185, USA}

\author{E. P. Nordberg}
\affiliation{Sandia National Laboratories, Albuquerque, New Mexico 87185, USA}
\affiliation{University of Wisconsin-Madison, Madison, Wisconsin 53706, USA}

\author{R. W. Young}
\affiliation{Sandia National Laboratories, Albuquerque, New Mexico 87185, USA}

\author{C. Borr\'{a}s Pinilla}
\affiliation{University of Oklahoma, Norman, Oklahoma 73019, USA}
\affiliation{Universidad Industrial de Santander-Colombia, Bucaramanga, Colombia}

\author{H. L. Stalford}
\affiliation{Sandia National Laboratories, Albuquerque, New Mexico 87185, USA}
\affiliation{University of Oklahoma, Norman, Oklahoma 73019, USA}

\author{G. A. Ten Eyck}
\affiliation{Sandia National Laboratories, Albuquerque, New Mexico 87185, USA}

\author{K. Eng}
\affiliation{Sandia National Laboratories, Albuquerque, New Mexico 87185, USA}

\author{K. D. Childs}
\affiliation{Sandia National Laboratories, Albuquerque, New Mexico 87185, USA}

\author{J. R. Wendt}
\affiliation{Sandia National Laboratories, Albuquerque, New Mexico 87185, USA}

\author{R. K. Grubbs}
\affiliation{Sandia National Laboratories, Albuquerque, New Mexico 87185, USA}

\author{J. Stevens}
\affiliation{Sandia National Laboratories, Albuquerque, New Mexico 87185, USA}

\author{M. P. Lilly}
\affiliation{Center for Integrated Nanotechnologies, Sandia National Laboratories, Albuquerque, New Mexico 87185, USA}

\author{M. A. Eriksson}
\affiliation{University of Wisconsin-Madison, Madison, Wisconsin 53706, USA}

\author{M. S. Carroll}
\affiliation{Sandia National Laboratories, Albuquerque, New Mexico 87185, USA}

\date{\today}

\begin{abstract}
We present transport measurements of a tunable silicon metal-oxide-semiconductor double quantum dot device with lateral geometry.  
Experimentally extracted gate-to-dot capacitances show that the device is largely symmetric under the gate voltages applied.  
Intriguingly, these gate voltages themselves are not symmetric.  
Comparison with numerical simulations indicates that the applied gate voltages serve to offset an intrinsic asymmetry in the physical device.
We also show a transition from a large single dot to two well isolated coupled dots, where the central gate of the device is used to controllably tune the interdot coupling.

\end{abstract}

\pacs{}

\maketitle 

Recent progress towards demonstrating electron spin-based quantum bits in semiconductor quantum dots \cite{Loss,Petta,Koppens} 
has lead to renewed interest in fabrication of quantum dot structures in silicon.  Silicon is a strong candidate 
due to a relatively small electronic spin-orbit coupling, low concentration of nuclear spins (a potential source of electron spin decoherence),
and the ability to leverage mature silicon fabrication technologies.

We report low-temperature transport measurements of a Si metal-oxide semiconductor (MOS) double quantum dot (DQD).
In contrast to previously reported measurements of DQD's in Si MOS structures \cite{Rokhinson,FujiwaraAPL2006,Kim,FujiwaraAPL2008,CQCT2009,YangPRB2009}, 
our device has a lateral gate geometry 
very similar to that used by Petta \textit{et al.} \cite{Petta} to demonstrate coherent manipulation of single electron spins.  
This gate design \cite{Ciorga} provides a high degree of tunability, 
allowing for independent control over individual dot occupation and tunnel barriers, as well as
the ability to use nearby constrictions to sense dot charge occupation \cite{NordbergAPL}.

\begin{figure*}
\includegraphics[width = 165mm]{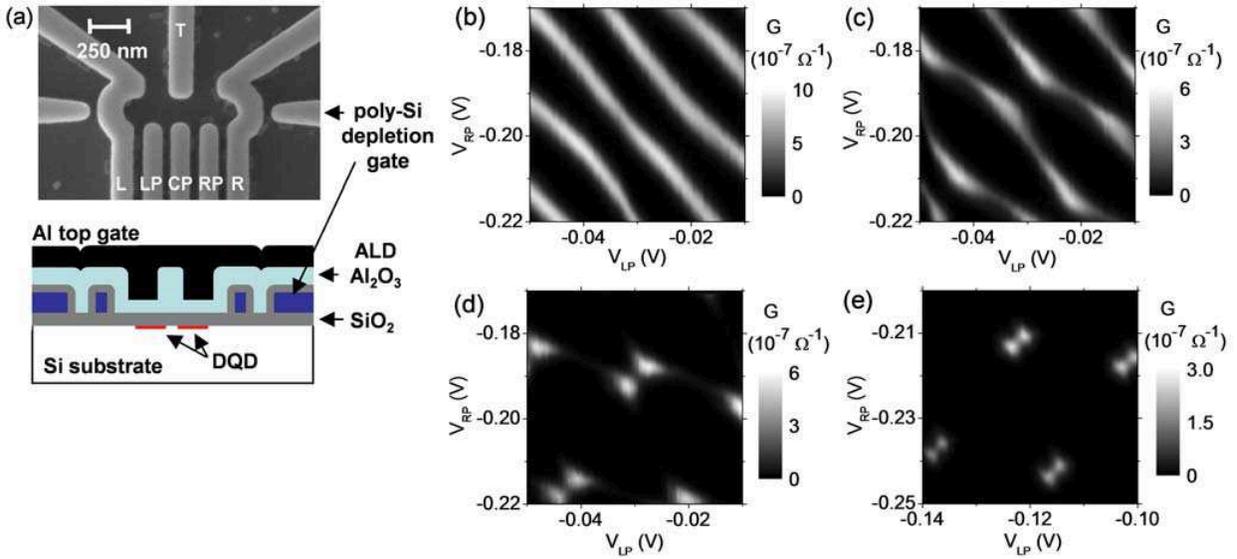}
\caption{\label{FIG. 1} (a) SEM image of partially processed device showing polysilicon gates and schematic of final device cross-section. 
(b) - (e) Conductance of double dot G versus \variable{V}{LP} and \variable{V}{RP}, showing transition from a large single-dot to
a well-defined double dot, at CP gate voltages (b) \variable{V}{CP} = -0.6 V, (c)  \variable{V}{CP} = -0.8 V, (d)  \variable{V}{CP} = -1.0 V, 
(e)  \variable{V}{CP} = -1.2 V, for \variable{V}{T} = -0.3 V,  \variable{V}{L} = 0 V, and  \variable{V}{R} = -2~V.}
\end{figure*}

Figure 1(a) shows a top-down scanning electron microscope (SEM) image of the partially processed device and a schematic of the 
final device cross section.  The upper Al top gate is used to accumulate carriers at the Si-SiO$_2$ interface.  The lower, patterned
polycrystalline silicon (polysilicon) gates are used to deplete carriers to define the DQD region.  The device is fabricated on a lightly p-type Si substrate
(2 - 20 $\Omega$ cm).  Initial fabrication steps (Ohmic contact formation to channel, gate oxide growth, and polysilicon deposition and
doping) are performed in a fully qualified CMOS facility.  Ohmic contacts (not shown in Fig.~1(a)) are formed by implantation of As, followed
by a 900 $^{\circ}$C, 15 min.~activation anneal.  Next, a 35 nm SiO$_2$ gate oxide is grown in dry O$_2$ at 900$^{\circ}$C, with a subsequent
N$_2$ anneal at 900$^{\circ}$C for 30 min., immediately followed by polysilicon deposition (200 nm) and doping.  The lower-level polysilicon
depletion gates are defined by e-beam lithography and dry etch and are 200 nm tall with a $\sim$100 nm linewidth for the narrowest features (see Fig.~1(a)).
After the polysilicon etch, the device is exposed to a thermal oxidation step which forms a 30 nm layer of SiO$_2$ on the surface of the polysilicon gates.
Finally, the polysilicon is further isolated from the global Al top gate by 60 nm of Al$_2$O$_3$ formed via atomic layer deposition (ALD).  
More details about the fabrication of this device and its operation in single-dot mode can be found in Ref. \onlinecite{NordbergPRB}.

The device conductance is experimentally determined via standard low-frequency lock-in measurements with an rms ac source-drain bias of 
10 - 50 \textmu{}V.  Unless otherwise noted, the Al top gate TG, T, L, and R (see Fig~1(a)) voltages are held constant at
 \variable{V}{TG} = 5 V,  \variable{V}{T} = -0.3 V,  \variable{V}{L} = 0 V, and  \variable{V}{R} = -2 V.
All measurements shown are performed in a dilution refrigerator with a temperature of $T \sim 20$ mK.  
We note that although the conductance of the device continues to evolve below a fridge temperature of 100 mK, we cannot be certain our dot electron temperature is
equal to our fridge temperature in this regime.  However, the precise value of the electron temperature should not affect the conclusions of this
letter.

Figures 1(b)-(e) show DQD conductance versus left and right plunger gate voltages \variable{V}{LP} and \variable{V}{RP}
at four different center plunger voltages \variable{V}{CP},
demonstrating the ability to tune the DQD from a highly-coupled regime, where the transport is reminiscent of that expected for a large, single
dot, to a weakly-coupled regime, where conduction can only take place at the so-called triple points.  As expected, the interdot coupling shows a strong dependence
on \variable{V}{CP}, whereas we find a nearly negligible dependence on \variable{V}{LP} and \variable{V}{RP}.

\begin{figure}
\includegraphics[width = 90mm]{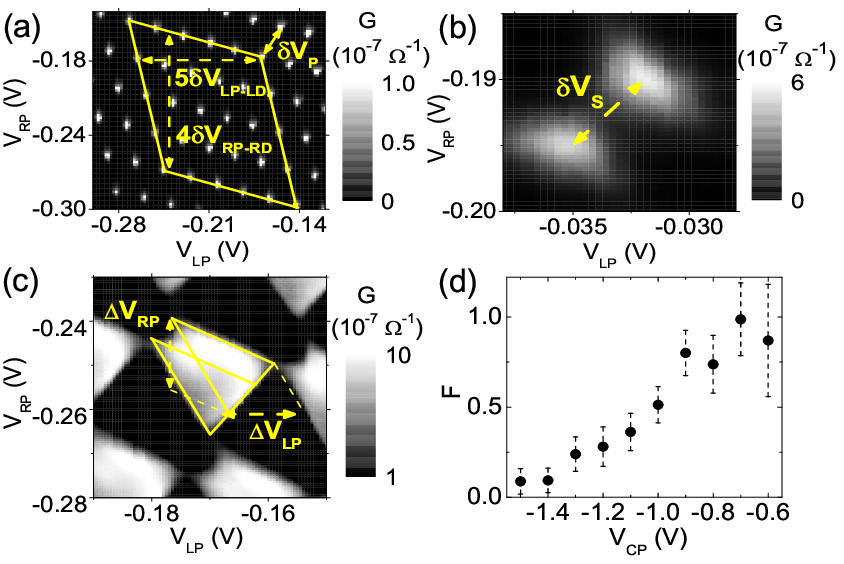}
\caption{\label{FIG. 2}  (a) DQD conductance $G$ vs. \variable{V}{LP} and \variable{V}{RP} at \variable{V}{CP} = -1.2 V, 
showing LP (RP) voltage required to change the occupation of the left dot 
 (right dot) by 5 (4) electrons: $5\mathrm{\delta} \variable{V}{LP-LD}$ ($4\mathrm{\delta} \variable{V}{RP-RD}$).
(b) Closer view of triple points at \variable{V}{CP} = -1.2 V for zero dc source drain bias.  
(c) Triple points at \variable{V}{CP} = -1.2 V for dc source-drain bias $\variable{V}{sd} = 0.5$ mV, showing dimensions of bias triangles.  
(d) Interdot coupling $F$ versus \variable{V}{CP} determined from plunger gate voltage separation of triple points.}
\end{figure}

In Fig.~2(a) we show DQD conductance versus \variable{V}{LP} and \variable{V}{RP} at \variable{V}{CP} = -1.2 V. 
The capacitance between the plunger gates and dots can be extracted from the dimensions shown in Fig.~2(a):
$\variable{C}{LP-LD} = e/\mathrm{\delta} \variable{V}{LP-LD} = 8.1$ aF, $\variable{C}{RP-RD} = e/\mathrm{\delta} \variable{V}{RP-RD} = 5.3$ aF.
Figure~2(c) shows bias triangles at a finite dc source-drain bias of 0.5 mV at \variable{V}{CP} = -1.2 V.   
Using $\variable{C}{LP-LD(RP-RD)} / \variable{C}{LD(RD)} = \variable{V}{sd} / \mathrm{\Delta} \variable{V}{LP(RP)}$ we find total dot capacitances 
$\variable{C}{LD} = 210$ aF and $\variable{C}{RD} = 170$ aF \cite{vanderWiel}.

Figure 2(b) shows a closer view of single pair of triple points at \variable{V}{CP} = -1.2 V and zero dc source-drain bias.
The distance between triple points in plunger gate voltage space $\mathrm{\delta} \variable{V}{S}$, as shown in Fig.~2(b), can be used
to estimate the interdot coupling.  In Fig.~2(d) we show the fractional splitting of the triple point $F$ versus CP gate voltage, 
where $F \equiv 2 \mathrm{\delta} \variable{V}{S} / \mathrm{\delta} \variable{V}{P}$ (see Fig.~2(a), (b)) is a measure of the interdot coupling, 
defined as the ratio of the diagonal separation between triple points to the separation between charge domains \cite{Westervelt,Halperin}.
The error bars are determined by the ability to visually resolve the position of the triple points on a honeycomb plot.   Figure~2(d) shows that sweeping CP
smoothly varies $F$ from nearly zero to one, demonstrating the ability to tune the DQD from weakly-coupled to a regime where the dots are fully merged
\cite{Westervelt,Simmons}.

Figure 3(a) shows a comparison between experimentally determined capacitances from the various gates to the dots and values determined via modeling.  
The values $\variable{C}{MEAS}$ in Fig.~3(a) are obtained from honeycomb plot dimensions from double dot transport, or from the gate voltage period of Coulomb blockade oscillations
in single dot transport.  For double dot transport, the gate voltages are \variable{V}{TG} = 5 V, \variable{V}{T} = -0.3 V, 
\variable{V}{L} = 0 V, \variable{V}{LP} = 0 V, \variable{V}{CP} = -1.2 V, \variable{V}{RP} = 0 V, \variable{V}{R} = -2 V.
Values marked with an asterisk indicate capacitances from single dot transport, obtained for the left dot by setting \variable{V}{R} = +1 V, or for the right dot
by using \variable{V}{L} = +1 V, while maintaining all other gate voltages the same as used for double dot transport.  Error bars are determined by variations in the charge transition period.

Figure 3(b) shows a contour plot of calculated electron density in the DQD region at gate voltages equal to those used for the experimental capacitances. 
The calculations are semiclassical (Thomas-Fermi approximation) and were performed using TCAD Sentaurus \cite{Synopsis}, a commercial device simulation package.
The capacitances \variable{C}{Sentaurus} are obtained by integrating the charge density over the entire left or right dot region and calculating the change in the integrated
dot charge due to a small change in gate voltage.   The boundaries used to separate the left dot from the right dot and the dot region from the leads are shown by the dashed lines in Fig.~3(b).

In Fig.~3(c), we show a model dot region (shaded area) used to calculate capacitances between the various gates and dots by treating the dot region
as a perfectly-conducting metallic sheet.  These capacitance calculations were performed using CFD-ACE+ \cite{CFD-ACE}, a finite-element modeling package.
The shape of the dot was chosen by starting with a region which approximately follows the contours defined by the gates, and then modifying the 
distance between the dot and various gates as to obtain good agreement with the experimental capacitances $\variable{C}{MEAS}$ listed in Fig.~3(a).

\begin{figure}
\includegraphics[width = 85mm]{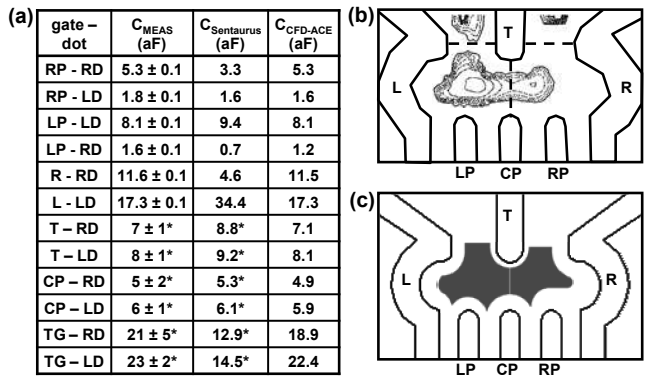}
\caption{\label{FIG. 3} (a) Table of capacitances between dots (right dot (RD) and left dot (LD)) and gates (as labeled in Fig.~1(a), comparing measured and calculated values.  
$\variable{C}{MEAS}$ are experimental values for the case \variable{V}{CP} = -1.2 V, $\variable{C}{Sentaurus}$ values are computed using a semiclassical model for dot electron density, 
and $\variable{C}{CFD-ACE}$ values are calculated by treating the dots as metallic regions defined by the shaded areas sketched in (b).  Asterisks indicate
capacitances from single-dot transport, obtained by setting \variable{V}{R} (\variable{V}{L})= +1 V for left (right) dot measurement.
(b) Calculated electron density in dot regions relative to polysilicon gates, where the dotted lines are contours of electron density
from 1 to $7 \times 10^{11}$ cm$^{-2}$ in $1 \times 10^{11}$ cm$^{-2}$ increments.  
The dashed lines define the boundary between the left and right dot regions and between the dot regions and the leads.
(c) Sketch of dot region used to obtain good agreement between experimental and calculated capacitances.  Shaded area represents dot region, which is treated as
a metallic (perfect conductor) region in order to calculate capacitances.}
\end{figure}

The agreement between the capacitances predicted by the Sentaurus calculation and those obtained from experiment is relatively poor.
As shown in Fig.~3(b), the Sentaurus calculation predicts an asymmetric dot region, resulting in a capacitance of $\variable{C}{R-RD} = 4.6$ aF from
the R gate to right dot, versus $\variable{C}{L-LD} = 34.4$ aF from the L gate to left dot.
This extreme asymmetry does not reflect transport in the actual device, which yields $\variable{C}{R-RD} = 11.6$ aF versus $\variable{C}{L-LD} = 17.3$ aF.
The dot regions shown in Fig.~3(c), which result in reasonable agreement between $\variable{C}{MEAS}$ and $\variable{C}{CFD-ACE}$,
predict a right dot that is smaller than the left dot, but not as dramatically so as suggested by the contours in Fig.~3(b). 

The disagreement between \variable{C}{MEAS} and \variable{C}{Sentaurus} suggests that the Sentaurus calculation fails to capture some aspect of the actual device,
such as unintentional imperfections that may be present in the experiment, but are not included in the calculation.
For example, a spatially varying distribution of charge in the dielectric regions or variations in film thickness (e.g. non-uniform Al$_2$O$_3$ deposition)
can lead to unintentional variations in electron density in the dot region.
However, the relative symmetry of the dot contours shown in Fig.~3(c) (as compared to Fig.~3(b)) suggests that by use of 
asymmetric gate voltages ($\variable{V}{R} = -2$ V versus $\variable{V}{L} = 0$ V) we are able to
partially compensate for this disorder.

In conclusion, we have demonstrated a gate-defined Si MOS double quantum dot with lateral geometry.
Comparison between experimental and calculated capacitances suggest the presence of disorder 
and that we are able to partially compensate for this disorder by adjustment of gate voltages, due to the tunability of our lateral geometry.
Our data also shows the ability to control the interdot coupling over a wide range of energies.
Similar device structures may allow for further study of DQDs in the few electron regime, which would be of interest for quantum computing applications.

\begin{acknowledgments}
This work was performed, in part, at the Center for Integrated Nanotechnologies, 
a U.S. DOE, Office of Basic Energy Sciences user facility, and was supported by the Laboratory Directed Research and Development program
at Sandia National Laboratories, a multi-program laboratory operated by Sandia Corporation, 
a wholly owned subsidiary Lockheed-Martin Company, for the U. S. Department of Energy's National Nuclear Security Administration under Contract No. DE-AC04-94AL85000.
\end{acknowledgments}

\end{document}